# Secure Cloud Computing through Homomorphic Encryption

[1]Maha TEBAA, [2]Said EL HAJII
[1]*Laboratory of Mathematics, Computer and Applications, University Mohammed V–Agdal, Faculty of Science, Rabat-Morocco, maha.tebaa@gmail.com*
[2]*Laboratory of Mathematics, Computer and Applications, University Mohammed V–Agdal, Faculty of Science, Rabat-Morocco, elhajji@fsr.ac.ma*

**Abstract**

*Go to the cloud, has always been the dream of man. Cloud Computing offers a number of benefits and services to its customers who pay the use of hardware and software resources (servers hosted in data centers, applications, software...) on demand which they can access via internet without the need of expensive computers or a large storage system capacity and without paying any equipment maintenance fees. But these cloud providers must provide guarantees on the protection of privacy and sensitive data stored in their data centers shared between multiple clients using the concept of virtualization.*

## 1. Introduction

Cloud Computing has emerged as an important paradigm that has attracted considerable attention in both industry and academia. Cloud Computing already existed under different names like "outsourcing" and "server hosting." But the poor performance of processors used, slow Internet connections and the exorbitant costs of the materials used, do not allow the use of services and storage spaces. However, recent advances in current technology (through virtualization) paved the way for these operations with faster processing.

Cloud Computing security challenges and it's also an issue to many researchers; first priority was to focus on security which is the biggest concern of organizations that are considering a move to the cloud. The use of cloud computing brings a lot of advantages including reduced costs, easy maintenance and reprovisioning of resources. The first real use of the concept of cloud computing was in 2002 by the company Amazon Web Services, when it leased its resources to companies during periods off celebrations (when there was no peak usage of its IT) on demand.

Many people use the cloud every day without knowing. For example in all versions of email (Gmail or Webmail) and access to the applications that are not physically installed on the local PC as Excel, Microsoft Word…this use is done thanks to Internet, but customers may not know the location of the servers that storing their emails and hosting the source code of the applications that they use. The services offered by the Cloud Computing providers, come from huge digital stations called Data-centers, using techniques based on virtualization [1]. The virtualization is all the technical material and/or software that can run on a single machine multiple operating systems and/or multiple applications, separately from each other, as if they were working on separate physical machines. Virtualization and consolidation can simplify the management of the server's park, by reducing the number of machines to be maintained by optimizing the use of resources and enabling high availability. But the adoption and the passage to the Cloud Computing applies only if the security is ensured. How to guaranty a better data security and also how can we keep the client private information confidential? There are two major questions that present a challenge to Cloud Computing providers

## 2. The Different Virtualization Techniques

The growth of the activity has irreparably the need to evolve the IT infrastructure. Adding a new servers for new applications at risk of under-use others. Administration costs are increasing and the structure loses flexibility and reliability. Among the reasons for adopting virtualization are server consolidation and infrastructure optimization, virtualization can significantly increase the rate of resource utilization by pooling common resources and leaving the scheme "application = server". Thanks to virtualization you can reduce the number of servers and the amount of hardware needed in the data-center. This is represented by lower real estate costs and the need for power and cooling, resulting in a net reduction of IT costs.
Increased flexibility and operational efficiency: Virtualization offers a new way of managing IT infrastructure and can help IT administrators spend less time on repetitive tasks such as provisioning, monitoring and maintenance. It also enables increased availability of applications, improved continuity of services, save and move safely entire virtual environments without interrupting services and improving the management and security of workstations: administer, deploy, manage and monitor closely the servers.

Virtualization techniques [2] are numerous, and the choice of the proper technique requires a detailed study of IT platform to virtualize.

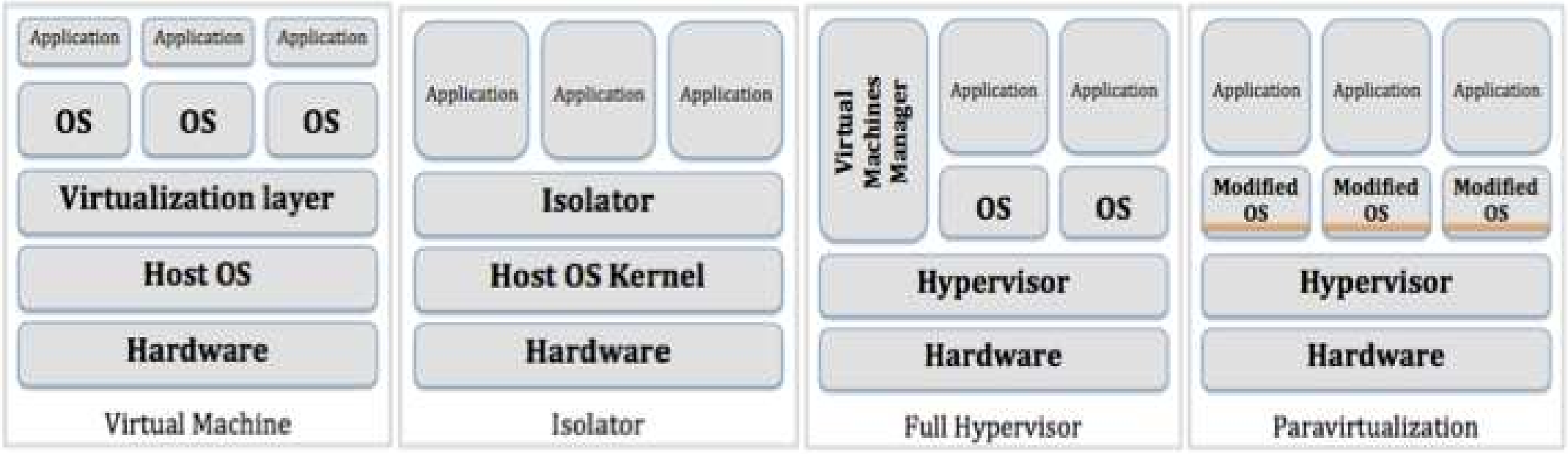

**Figure 1.** Various Types of Virtualization

***A. Virtual Machine:***
A Virtual Machine is software that runs on the operating system (OS) host. This software allows you to run one or more guest OS, each of which has access at virtualized hardware but it thinks interact directly with the physical hardware, this solution allows to isolate the guest operating systems but its a weak point is high consumption of resources (reduced performance of I/O and limitation of CPU performance).

***B. Isolator***:
An Isolator is software to isolate the execution of applications in the different performance's areas. The isolator allows running several times the same application in a multi- instance mode (multiple running instances) even if it was not designed for that.

***C. FullHypervisor:***
A Hypervisor is software that runs directly on a hardware platform, this platform is considered as a tool to control the operating system. Therefore, a secondary operating system can be run over the material. The hypervisor is a lightweight and optimized host kernel to execute guest operating systems that are adapted to a specific architecture.

***D. Paravirtualization:***
The Paravirtualization based on Hypervisor system changes virtualized OS in order to be adapted and optimized for this use. It allows virtual machines to communicate with each other and especially with the parent machine through a software bus. The paravirtualized machines not exploit the emulated drivers but directly the physical drivers even for managing I/O.

## 3. Characteristics of Cloud Computing Services

Definition [3]: By cloud computing we mean: The Information Technology (IT) model for computing, which is composed of all the IT components (hardware, software, networking, and services) that are necessary to enable development and delivery of cloud services via the Internet or a private network. Among the characteristics of Cloud [4], we quote:

- **Elasticity:** it is one of the most essential characteristics of our vision of Cloud. It defines the ability of a given infrastructure to dynamically adapt to a scale.

- **Ability to adapt:** Cloud must provide a set of automatization allowing it self-management. Its administration should require a minimum human intervention.

- **Quality of Service:** is another key aspect of Cloud, using metrics such as time response, the number of operations in a second; the service provides guarantees to its users. It no longer belongs to the user having to decide what resources to deploy but rather to define terminals that the service should meet. Cloud adapts to ensure its terminals.

- **High Availability:** playing on replicated data in different data centers, the Cloud must provide reliable, not sensitive to the failure of an instance or a data center.

- **Cost reduction:** Pay Per Use, means that the useonly pays for the service based on its utilization.

- **Ecological approach:** the allocation of resources to the strict necessity to reduce the energy consumption of IT parks. Beyond the economic aspect, these reductions allow the ecological energy reduction footprint of the company.

A company may opt for different structures of clouds [5] tailored to specific needs. We quote:

- ***Public Cloud:*** As the name suggests, it is to share with the "public" (user at large) an infrastructure that belongs to a cloud provider, which leases its services to companies on demand. Its main role is to host applications, web in general, only accessible via the Internet, so it is an optimal pooling of resources based on the creation of a multitude of execution environment on a same platform. The bill for this type of cloud doesn't contain subscription or commissioning fees, the actual consumption is paid each end-of-month.

- ***Private Cloud:*** Is to transform the internal infrastructure of a computer system through virtualization technologies, providing services and resources to clients on demand. These services are hosted by the client company or by the cloud provider (with a VPN connection). We distinguish: internal private clouds, used by the company to meet its own needs, they are administered internally by the company itself, there are also external private clouds, their directions are assigned to an outside provider.

- ***Community Cloud:*** The infrastructure is shared by several organizations that have common interests (e.g. security requirements, compliance considerations...). As the Private Cloud, it can be managed by the organizations themselves or by a third party.

- ***Hybrid Cloud:*** Is to coexist and communicate a private cloud and public cloud. The infrastructure consists of two or more clouds (private, Community or Public). The hybrid is often used to encompass the peak time charges as with the public, remaining linked to private Cloud. Both infrastructure therefore communicate and form a hybrid cloud, then it is a way to combine the benefits of both platforms.

The NIST (National Institute of Standards and Technology), counts three service models of cloud computing [6]:

- ***Cloud Software as a Service (SaaS):*** This is the application part. The applications are accessible via the Internet, and offered as a subscription with a payment to use. This is the visible part of the cloud to the end user, who does not need to install the application on his computer, but it can be accessed via various interfaces, thin client, Web, and mobile browser... The company does not manage or control the underlying cloud infrastructure including network, servers, operating systems, storage, or even individual application capabilities, with the possible exception of limited user-specific application configuration settings.
- ***Cloud Platform as a Service (PaaS):*** The user can deploy their own cloud infrastructure applications, using the programming language, libraries, services, and tools supported by the provider. The consumer does not manage or control the underlying cloud infrastructure including network, servers, operating systems, or storage, but has control over the deployed applications and possibly configuration settings for the application-hosting environment.
- ***Cloud Infrastructure as a Service (IaaS):*** It is the physical infrastructure rented on demand: storage, virtual machines, OS...The consumer does not manage or control the underlying cloud infrastructure but has control over operating systems, storage, and deployed applications; and possibly limited control of select networking components (e.g., host firewalls).

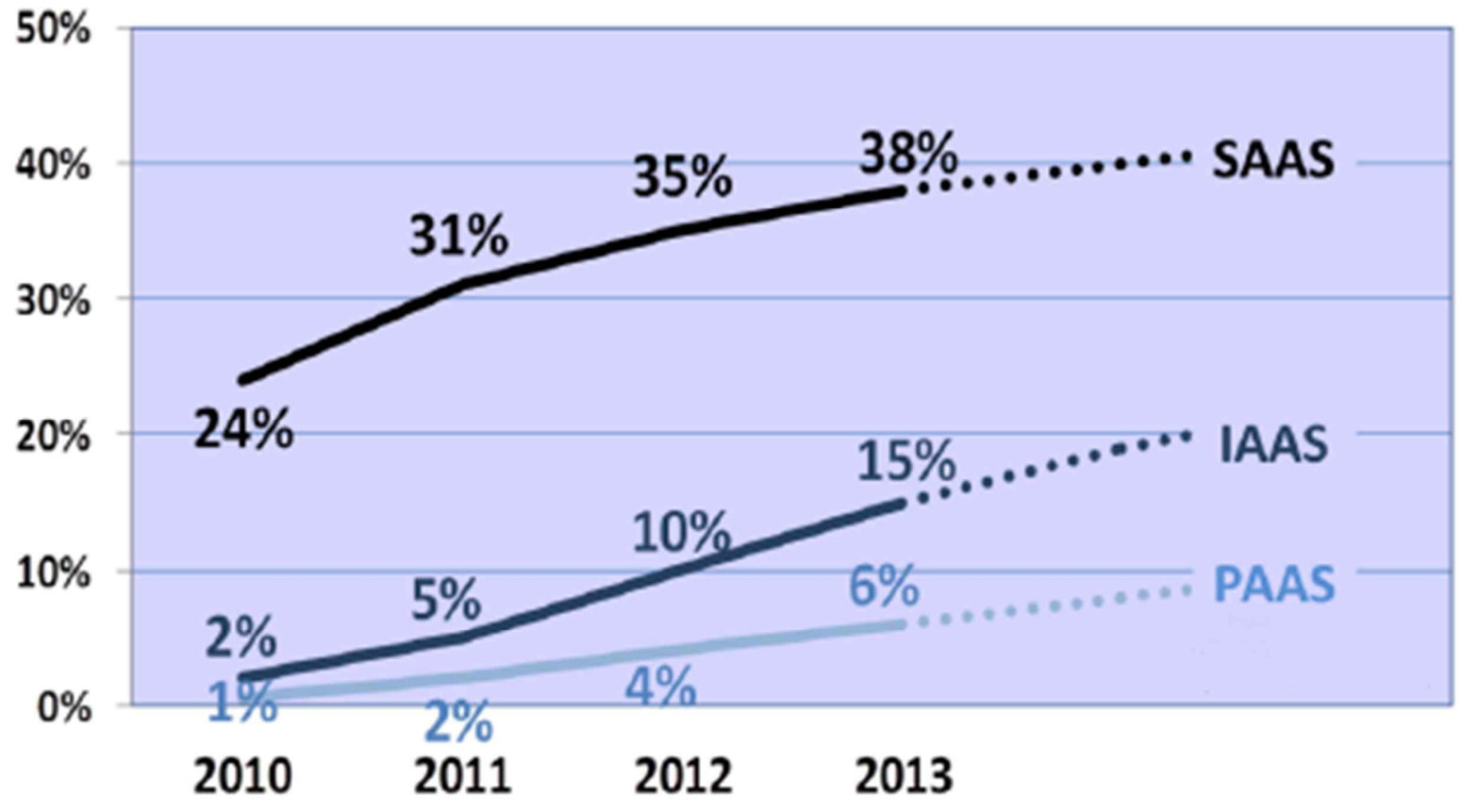

***Source: MARKESS International***

**Figure 2.** Evolution of Cloud Computing Adoption by the French Organizations, 2010-2013 [7]

## 4. Security Issues for cloud computing

The definition of cloud computing that we mentioned in the previous section doesn't mention any security notion of the data stored in the Cloud Computing even being a recent definition. Therefore we understand that the Cloud Computing is lacking security, confidentiality and visibility. To Provide Infrastructure (IaaS), Platform Service (PaaS) or Software (SaaS) as a Service is not sufficient if the Cloud provider does not guaranty a better security and confidentiality of customer's data.

By convention, we consider as Cloud Computing any treatment or storage of personal or professional information which are realized outside the concerned structure (i.e outside the company), to secure the Cloud means secure the treatments (calculations) and storage (databases hosted by the Cloud provider).

Cloud providers such as IBM, Google and Amazon use the virtualization on their Cloud platform and on the same server can coexist a virtualized storage and treatment space that belong to concurrent enterprises.

The aspect of security and confidentiality must intervene to protect the data from each of the enterprises. Secure storage and treatment of data requires using a modern aspect of cryptography that has the criteria for treatment such as, the necessary time to respond to any request sent from the client and the size of an encrypted data which will be stored on the Cloud server.
Transfer the processing of your data to a third party; it is also transferring some of the responsibility associated with their security and compliance. It is not surprising that security professionals are nervous. Therefore, it is essential that you fully trust in your cloud provider. The advantages of cloud computing include reduced costs, easy maintenance and re- provisioning of resources, and thereby increased profits. Our proposal is to encrypt data before sending it to the cloud provider, but to execute the calculations the data should be decrypted every time we need to work on it. Until now it was impossible to encrypt data and to trust a third party to keep them safe and able to perform distant calculations on them. So to allow the Cloud provider to perform the operations on encrypted data without decrypting them requires using the cryptosystems based on Homomorphic Encryption [8].

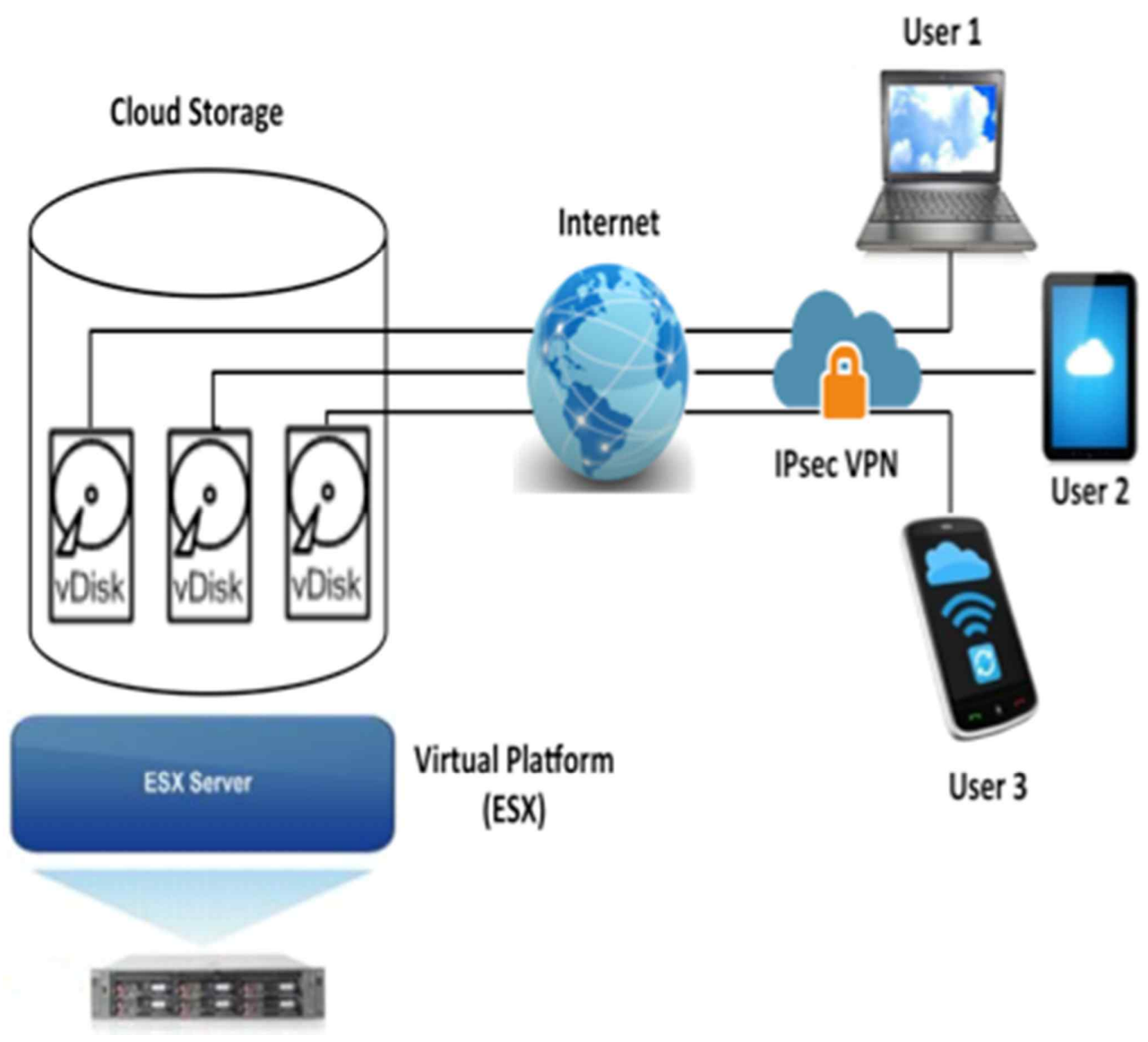

**Figure 3.** A scheme representing the basic architecture of Cloud Computing

## 5. Homomorphic Encryption Applied to Cloud Computing Security

When the data transferred to the Cloud we use standard encryption methods to secure the operations and the storage of the data. Our basic concept was to encrypt the data before send it to the Cloud provider. But the last one needs to decrypt data at every operation. The client will need to provide the private key to the server (Cloud provider) to decrypt data before execute the calculations required, which might affect the confidentiality and privacy of data stored in the Cloud.
In this paper we are proposing an application of a method to execute operations on encrypted data without decrypting them, which will provide the same results after calculations as if we have worked directly on the raw data.

Homomorphic Encryption systems are used to perform operations on encrypted data without knowing the private key (without decryption), the client is the only holder of the secret key. When we decrypt the result of any operation, it is the same as if we had carried out the calculation on the raw data.
Definition [9]: An encryption is homomorphic, if: from Enc(a) and Enc(b) it is possible to compute Enc(f (a, b)), where f can be: $+$, $\times$, $\oplus$ and without using the private key. Among the Homomorphic encryption we distinguish, according to the operations that allows to assess on raw data, the additive Homomorphic encryption (only additions of the raw data) is the Pailler [10] and Goldwasser-Micalli [11] cryptosystems, and the multiplicative Homomorphic encryption (only products on raw data) is the RSA [12] and El Gamal [13] cryptosystems.

- $E_k$ is an encryption algorithm with key k.
- $D_k$ is a decryption algorithm.

$D_k$ ($E_k$ (n) × $E_k$ (m)) = n×m OR Enc (x⊗y) = Enc(x) ⊗ Enc(y)
$D_L$ ($E_L$ (n) × $E_L$ (m)) = n+m OR Enc (x⊕y) = Enc(x) ⊗ Enc(y)

The first property is called additive homomorphic encryption, and the second is multiplicative homomorphic encryption. An algorithm is fully homomorphic if both properties are satisfied simultaneously.

*A. Multiplicative Homomorphic Encryption (RSA cryptosystem):*

Let $n = pq$ where $p$ and $q$ are primes. Pick $a$ and $b$ such that $ab \equiv 1 \pmod{\phi(n)}$. $n$ and $b$ are public while $p$, $q$ and $a$ are private.

$$e_K(x) = x^b \bmod n$$

$$d_K(y) = y^a \bmod n$$

The Homomorphism: Suppose $x_1$ and $x_2$ are plaintexts. Then,

$e_k(x_1)\, e_k(x_2) = x_1^b x_2^b \bmod n = (x_1 x_2)^b \bmod n = e_k (x_1 x_2)$

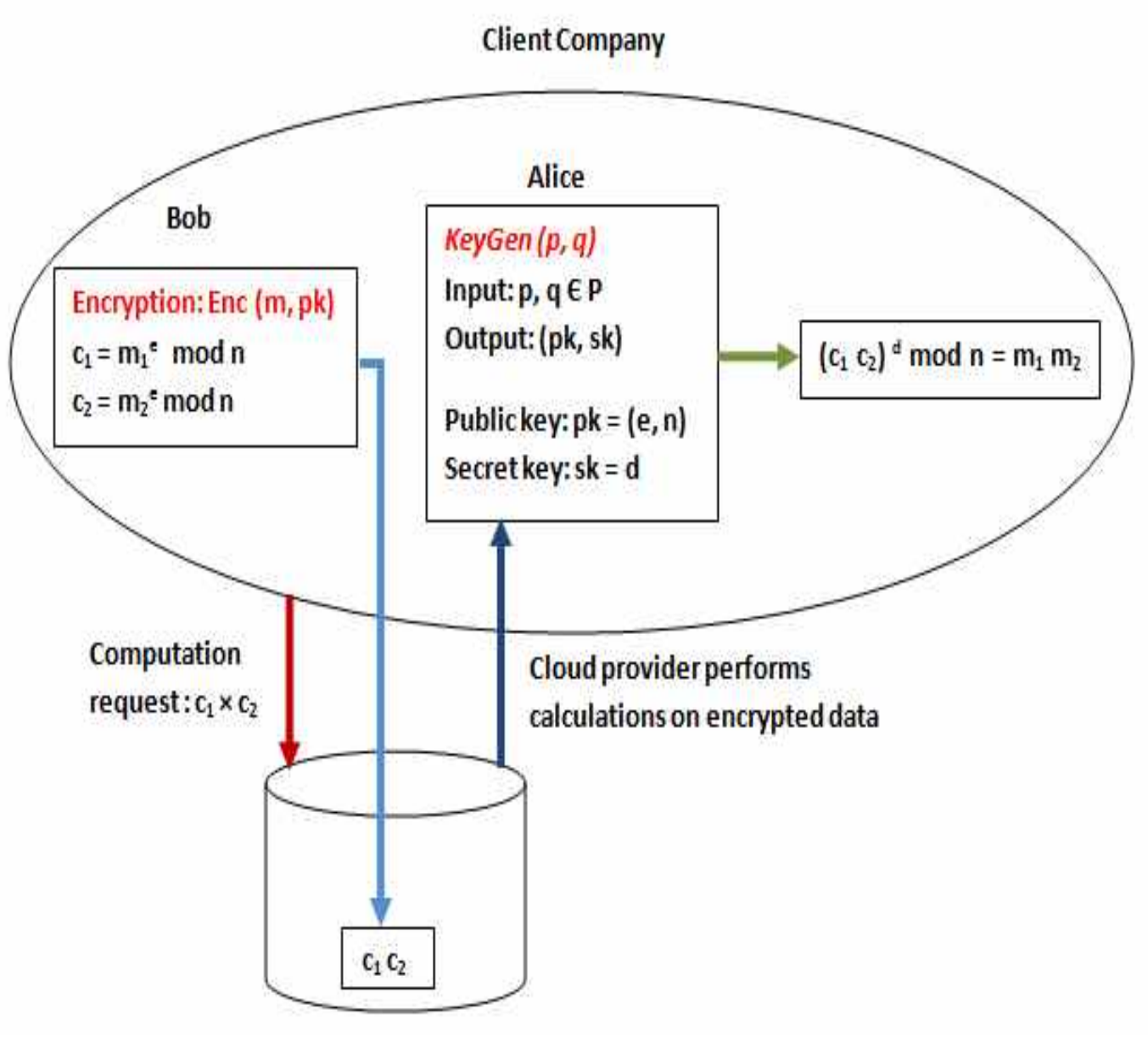

**Figure 4.** Multiplicative Homomorphic Encryption Applied to Cloud Computing

*B. Additive Homomorphic Encryption (Paillier Cryptosystem):*

Pick two large primes $p$ and $q$ and let $n = pq$. Let $\lambda$ denote the Carmichael function, that is, $\lambda(n) = \text{lcm}(p-1, q-1)$. Pick random $g \in \mathbb{Z}^*_{n^2}$ such that $L(g^\lambda \bmod n^2)$ is invertible modulo $n$ (where $L(u) = \frac{u-1}{n}$). $n$ and $g$ are public; $p$ and $q$ (or $\lambda$) are private. For plaintext $x$ and resulting ciphertext $y$, select a random $r \in \mathbb{Z}^*_n$. Then,

$$e_K(x, r) = g^x \, r^n \bmod n^2$$

$$d_K(y) = \frac{L(y^\lambda \bmod n^2)}{L(g^\lambda \bmod n^2)} \bmod n$$

The Homomorphism: Suppose x1 and x2 are plaintexts. Then,

$$\begin{aligned} e_k(x_1, r_1)\, e_k(x_2, r_2) &= g^{x1}\, r_1^{\,n}.\; g^{x2}\, r_2^{\,n} \bmod n^2 \\ &= g^{x1+x2}\, (r_1\, r_2)^n \bmod n^2 \\ &= e_k(x_{1+}\, x_2, r_1 r_2) \end{aligned}$$

To perform addition and multiplication on encrypted data stored in the cloud provider, the client must have two different key generators (one for RSA and one for Paillier). We present in what follows the El Gamal cryptosystem that is basically a multiplicative homomorphic cryptosystem but by modifying coding mode we can make it additive.

*C. El Gamal Cryptosystem:*

Let $p$ be a prime and pick $\alpha \in \mathbb{Z}^*_p$ such that $\alpha$ is a generator of $\mathbb{Z}^*_p$. Pick $a$ and $\beta$ such that $\beta \equiv \alpha^a \pmod p$. $p$, $\alpha$ and $\beta$ are public; $a$ is private. Let $r \in \mathbb{Z}_{p-1}$ be a secret random number. Then,

$$e_K(x, r) = (\alpha^r \bmod p, x\beta^r \bmod p)$$

El Gamal Cryptosystem performs multiplicative homomorphic encryption propriety:
Let $x_1$ and $x_2$ be plaintexts. Then,

$$\begin{aligned} e_k(x_1, r_1)\, e_k(x_2, r_2) &= (\alpha^{r1} \bmod p,\, x_1\, \beta^{r1} \bmod p)\, (\alpha^{r2} \bmod p,\, x_2\, \beta^{r2} \bmod p) \\ &= (\alpha^{r1+r2} \bmod p,\, (x_1\, x_2)\, \beta^{r1+r2} \bmod p \\ &= e_k(x_1 x_2, r_1 r_2) \end{aligned}$$

If we put the plaintext in the exponent, we get:

$$e_k(x, r) = (\alpha^r \bmod p,\, \alpha^x\, \beta^r \bmod p)$$

Then the homomorphism is additive:

$$\begin{aligned} e_k(x_1, r_1)\, e_k(x_2, r_2) &= (\alpha^{r1} \bmod p,\, \alpha^{x1}\, \beta^{r1} \bmod p)\, (\alpha^{r2} \bmod p,\, \alpha^{x2}\, \beta^{r2} \bmod p) \\ &= (\alpha^{r1+r2} \bmod p,\, \alpha^{x1+x2}\, \beta^{r1+r2}) \bmod p \\ &= e_k(x_{1+}x_2, r_{2+}r_2) \end{aligned}$$

*D. Fully Homomorphic Encryption [14]:*

For all types of calculation on the data stored in the cloud, we must opt for the fully Homomorphic encryption which is able to execute all types of operations on encrypted data without decryption.In 2009 Craig Gentry of IBM has proposed the first encryption system "fully homomorphic" that evaluates an arbitrary number of additions and multiplications and thus calculate any type of function on encrypted data. The application of fully Homomorphic encryption is an important stone in Cloud Computing security, more generally, we could outsource the calculations on confidential data to the Cloud server, keeping the secret key that can decrypt the result of calculation.

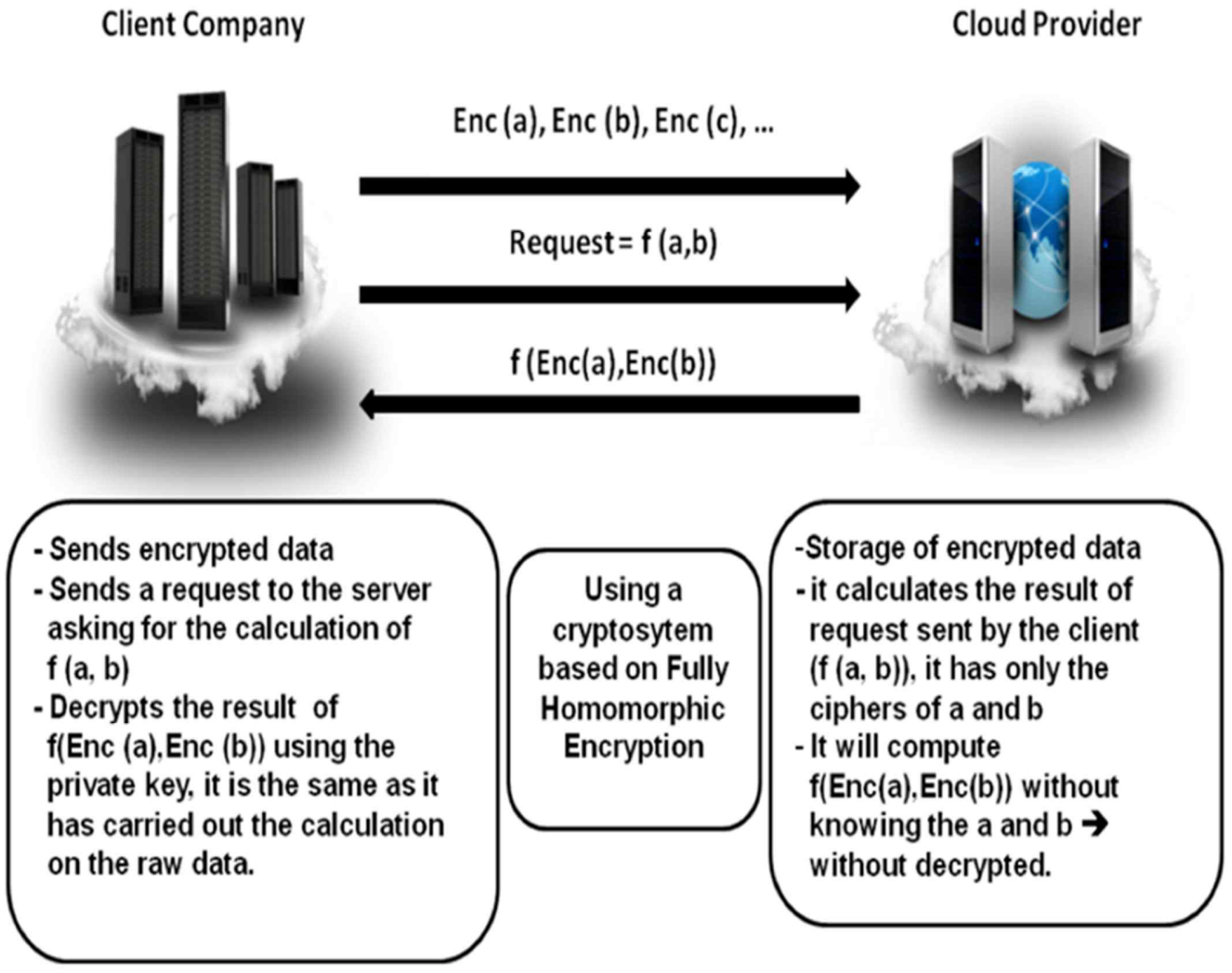

**Figure 5**. Fully Homomorphic Encryption applied to the Cloud Computing

## 6. Results and Discussions

In the table below we compare different Homomorphic Encryption cryptosystems according to the following characteristics:

- Homomorphic Encryption type.
- The respect for privacy of sensitive data.
- If security is applied at the provider of cloud or on customer.
- Who use the encryption and decryption keys?

In 2009 Craig Gentry of IBM has proposed the first encryption system "fully homomorphic" that evaluates an arbitrary number of additions and multiplications and thus calculate any type of function on encrypted data. The cryptosystem of Boneh-Goh-Nissim [15] recognizes a limitation stems from an error term that increases with each operation. Once the error tolerance is exceeded, the result cannot be decrypted.

| | Homomorphic Encryption Cryptosystems | | | | | |
|---|---|---|---|---|---|---|
| **Characteristics** | **RSA** | **Paillier** | **El Gamal** | **Goldwasser-Micali** | **Boneh-Goh-Nissim** | **Gentry** |
| **Platform** | Cloud Computing | Cloud Computing | Cloud Computing | Cloud Computing | Cloud Computing | Cloud Computing |
| **Homomorphic Encryption type** | Multiplicative | Additive | Multiplicative | Additive, but it can encrypt only a single bit | Unlimited number of additions but only one multiplication | Fully |
| **Privacy of data** | Is ensured in communication and storage processes | Is ensured in communication and storage processes | Is ensured in communication and storage processes | Is ensured in communication and storage processes | Is ensured in communication and storage processes | Is ensured in communication and storage processes |
| **Security applied to** | Cloud Provider Server | Cloud Provider Server | Cloud Provider Server | Cloud Provider Server | Cloud Provider Server | Cloud Provider Server |
| **Keys Used by** | The client (Different keys are used for encryption and decryption) | The client (Different keys are used for encryption and decryption) | The client (Different keys are used for encryption and decryption) | The client (Different keys are used for encryption and decryption) | The client (Different keys are used for encryption and decryption) | The client (Different keys are used for encryption and decryption) |

***Table 1.*** *Characteristics of existing Homomorphic Encryption Cryptosystems*

## 7. Conclusions

The Security of Cloud Computing based on fully Homomorphic Encryption is a new concept of security which is enable to provide the results of calculations on encrypted data without knowing the raw entries on which the calculation was carried out respecting the confidentiality of data.

This paper analyzes the application of different Homomorphic Encryption cryptosystems (RSA, Paillier, El Gamal, Goldwasser-Micali, Boneh-Goh-Nissim and Gentry) on a Cloud Computing platform. They are compared based on four characteristics; Homomorphic Encryption type, Privacy of data, Security applied to and keys used.

In future, we are going to analyse the behaviour of Homomorphic Encryption cryptosystems compared to the length of the public key and the time of the treatment of the request by the Cloud provider depending on the size of the encrypted messages.